\documentclass[
preprintnumbers,amsmath,amssymb]{revtex4}
\usepackage{graphicx}
\usepackage{dcolumn}
\usepackage{bm}

\usepackage{graphicx}
\usepackage{epsfig}

\begin{document} 
 \title{Integrable equations with Ermakov-Pinney nonlinearities and Chiellini damping}

 \author{Stefan C. Mancas}
\email{mancass@erau.edu  (phone: 386-226-7749, fax: 386-226-6269)}

\affiliation{Department of Mathematics, Embry-Riddle Aeronautical University,\\ Daytona Beach, FL. 32114-3900, U.S.A.}
\author{Haret C. Rosu}
\email{hcr@ipicyt.edu.mx  (phone: 444-834-2000, fax: 444-834-2010)}
\affiliation{IPICyT, Instituto Potosino de Investigacion Cientifica y Tecnologica,\\
Apdo. Postal 3-74 Tangamanga, 78231 San Luis Potos\'{\i}, S.L.P., Mexico\\}

\date{3 March 2015}

\begin{abstract}
\noindent %
We introduce a special type of dissipative Ermakov-Pinney equations of the form $v_{\zeta \zeta}+g(v)v_{\zeta}+h(v)=0$, where $h(v)=h_0(v)+cv^{-3}$ and the nonlinear dissipation $g(v)$ is based on the corresponding Chiellini integrable Abel equation.
When $h_0(v)$ is a linear function, $h_0(v)=\lambda^2v$, general solutions are obtained following the Abel equation route. Based on particular solutions, we also provide general solutions containing a factor with the phase of the Milne type. In addition, the same kinds of general solutions are constructed for the cases of higher-order Reid nonlinearities. The Chiellini dissipative function is actually a dissipation-gain function because it can be negative on some intervals.
We also examine the nonlinear case $h_0(v)=\Omega_0^2(v-v^2)$ and show that it leads to an integrable hyperelliptic case.\\ 


\medskip

\noindent {\em Highlights:}\\

$\bullet$  New type of dissipative Ermakov-Pinney equations is integrable.\\

$\bullet$  Nonlinear dissipation-gain introduced through the Chiellini integrable Abel equation.\\

$\bullet$  Higher-order Reid nonlinearities are also considered.\\

$\bullet$  Ermakov invariant is used in the solution method.

\end{abstract}

\begin{center}
Appl. Math. Comp. 259 (2015) 1-11
\end{center}

\maketitle

\newpage
\section{Introduction} \label{sec-Intro}
The nonlinear non-dissipative Ermakov-Pinney (EP) equations are known to have profound connections with the linear equations of identical operatorial form without the inverse cubic nonlinearity and because of this they are considered as an example of `nonlinearity from linearity' \cite{KD}. Leach and Andriopoulos \cite{LA} provide a historical overview and the fundamental importance of the EP equation for parametric oscillators, both classical and quantum-mechanical, with their vast application reaches is well established in the literature.
Recent works on the connections of the Ermakov systems with the nonlinear superposition principle belong to Cari\~nena and collaborators \cite{c1,c2}, where the reader can also find more references.
 The importance of Ermakov equations stems from the fact that they can occur in many research areas, such as in modeling the propagation of laser beams in nonlinear optics \cite{gon}, magneto-gas dynamics \cite{rs}, the mean field dynamics of pancake-shaped Bose-Einstein condensates \cite{herring}, and cosmology \cite{ro,hl,pe,daw}, to cite just a few.

 \medskip

On the other hand, the dissipative case and possible extensions, in spite of potential applications, are much less studied, deserving more attention \cite{c3}.
This gave motivation for this work in which we present, as main results, two integrable cases of nonlinear differential equations of the following {\em nonlinear} dissipative form
 \begin{equation}\label{ab-h}
 v_{\zeta \zeta}+g(v)v_\zeta+h(v)=0~,
 \end{equation}
 where
 \begin{equation}\label{ab-h1}
 h(v)= h_0(v)+c v^{-3}~.
 \end{equation}
 Regarding the function $h_0(v)$, we will consider in full detail the linear case $h_0(v)=\lambda^2v$ corresponding to a constant frequency $\omega_0^2=\lambda^2$, which makes (\ref{ab-h}) to have the simplest Ermakov-Pinney type format if one ignores the nonlinear dissipation $g(v)$. It is precisely the nonlinear dissipative ingredient which in general makes this equation nonintegrable. However, we will show here that this type of EP equation is integrable in the special case in which the nonlinear dissipation $g(v)$ is obtained from $h(v)$ through the Chiellini integrability condition of the Abel equation corresponding to (\ref{ab-h}).
We obtain the solution using the corresponding integrable Abel equation, and also we give a theorem for obtaining the general solution if a particular solution is known. In the latter case, the phase of the solution is of the Milne type \cite{Milne} and the Ermakov invariant for a pair of {\em nonlinear dissipative} EP equations of the type \eqref{ab-h} with different nonlinearity parameters $b$ and $c$ is used in the derivation. Furthermore, the same type of solutions are obtained for higher-order (Reid) negative power nonlinearities.
In the context of dissipative equations, we recall that for the general case of Ermakov equations with a {\em linear} dissipative term, it is known that one has to resort on numerical methods because there are no Lie symmetries and reductions to simpler forms are useful only in particular cases \cite{haas-10}.
 We also investigate the nonlinear case $h_0(v)=\Omega_0^2(v-v^2)$, or equivalently the frequency case $\omega_0^2(v)=\Omega_0^2(1-v)$, with the Chiellini dissipation, and show that it is also integrable.

\medskip

 The paper is structured as follows. We start by discussing briefly the basic properties of the simplest nondissipative EP equation corresponding to the constant frequency case $\omega_0^2=\lambda^2$, as mentioned above.
 We next move to the special EP equation with dissipation determined by Chiellini's integrability condition for the corresponding Abel equation of the first kind \cite{prev}, and the general solution is obtained through the usage of the Ermakov invariant of a pair of such dissipative EP equations. An application to the Courant-Snyder dynamics is included. The method is also applied to Chiellini-dissipative Ermakov equations with Reid (higher-order Ermakov) nonlinearities for which we also provide the general solutions. The last section is concerned with the nonlinear case $h_0(v)=\Omega_0^2(v-v^2)$ endowed with Chiellini's dissipation, followed by conclusions and an appendix in which we present a simplified case.

\section{The case ${\bf h_0(v)=\lambda^2v}$}\label{S2}

\subsection{Solutions of the simplest EP equation}\label{ssep}

Starting with the simple linear parametric oscillator equation,
\begin{equation}\label{a9}
u_{\zeta\zeta} +\omega^2(\zeta)u=0,
\end{equation}
it is a well-known fact that one can use two given linear independent solutions, $u_1$, and $u_2$,
to build a particular solution of the corresponding EP equation
\begin{equation}\label{a10}
v_{\zeta\zeta}+\omega^2(\zeta)v+cv^{-3}=0
\end{equation}
by means of Pinney's formula \cite{P}
\begin{equation}\label{a11}
v(\zeta)=\sqrt{u_1^2-\frac{cu_2^2}{W^2}}~,
\end{equation}
where $W$ is the Wronskian of the two solutions $u_1$, and $u_2$. Moreover, the general solution can be written
$$
v_g(\zeta)=\sqrt{\alpha_1 u_1^2+\alpha_2 u_2^2+2\alpha_3 u_1u_2}~,
$$
with the three constants constrained by the condition $\alpha_1\alpha_2-\alpha_3^2=-c/W^2$.

\medskip

Let us take the simplest case, i.e., $\omega^2(\zeta)=\lambda^2$, a constant:
\begin{equation}\label{a12}
u_{\zeta\zeta} +\lambda^2 u=0~.
\end{equation}

Then the corresponding EP equation reads
\begin{equation}\label{a13}
v_{\zeta\zeta}+\lambda^2v+cv^{-3}=0~.
\end{equation}
This equation will be called the simplest EP (SEP) equation henceforth.
From (\ref{a11}), the SEP particular solutions can be written immediately in the form
\begin{eqnarray}\label{a14}
\begin{array}{ll}
v_{-}(\zeta)=\sqrt{1-\Big(\frac{c}{\tilde {\lambda}^2}-1\Big)\,{\rm sinh}^2\tilde{\lambda}\zeta} ~, &  \, \lambda^2=-\tilde{\lambda}^2<0~,\\
\\
v_{0}(\zeta)=\sqrt{1-c\zeta^2}~, &  \, \lambda^2=0~,\\
\\
v_{+}(\zeta)=\sqrt{1-\Big(\frac{c}{\lambda^2}+1\Big)\sin^2\lambda\zeta}~, &   \, \lambda^2>0~.
\end{array} 
\end{eqnarray}

\medskip

\subsection{Chiellini dissipative SEP (CD-SEP) equations}\label{S4}

Noticing that the SEP equation (\ref{a13}) can be also written in the form
\begin{equation}\label{a13bis}
v_{\zeta\zeta}+h(v)=0, \qquad h(v)=\lambda^2v+cv^{-3}~,
\end{equation}
we build now CD-SEP equations, i.e., dissipative SEP equations with Chiellini-type damping as equations of the following format
 \begin{equation}\label{v-heq}
 v_{\zeta \zeta}+g(v)v_{\zeta}+h(v)=0~,
 \end{equation}
where $h(v)$ is as given in (\ref{a13bis}) and the nonlinear damping term $g(v)$ is obtained from $h(v)$ by means of Chiellini's integrability condition \cite{prev}
\begin{equation}\label{Chiell}
\frac{d}{dv}\left(\frac{h(v)}{g(v)}\right)=kg(v),\qquad \qquad k, \, {\rm a\,\, real\,\, constant}
\end{equation}
for the Abel equation of the first kind
\begin{equation}\label{Abel2}
\frac{dy}{dv}=g(v)y^2+h(v)y^3~.
\end{equation}
From (\ref{Chiell}) one easily gets
\begin{equation}\label{a19}
g(v)=\frac{\lambda^2v^2+cv^{-2}}{\sqrt{k \lambda^2 v^4+c_1 v^2-kc}}~,
\end{equation}
where $c_1$ is an integration constant.
On the other hand, the solution of Abel's equation \eqref{Abel2} is given by $y(v)=\frac{g(v)}{h(v)}|_{k=-2}$ as one can check by direct substitution 
and usage of (\ref{Chiell}), see also \cite{prev}. Thus, for CD-SEP equations, one gets the following Abel solution
\begin{equation}\label{a20}
y(v)=\frac{v}{\sqrt{-2 \lambda^2 v^4+c_1 v^2+2c}}~.
\end{equation}
Furthermore, using the substitution $v_\zeta\equiv 1/y(v)$ between \eqref{v-heq} and \eqref{Abel2} in \eqref{a20} and integrating, one has
\begin{equation}\label{a21}
\zeta-\zeta_0=\int \frac{vdv}{\sqrt{-2\lambda^2v^4+c_1v^2+2c}}~.
\end{equation}
We then obtain the following general solutions for the CD-SEP equations:

\begin{align}
v^-_{\tilde{\Lambda}}(\zeta)&= \left\{ \begin{array}{ll}
\frac{1}{2\tilde\lambda}\sqrt{-c_1+\sqrt{-\tilde{\Lambda}}\sinh{\big(2 \sqrt{2}\tilde\lambda(\zeta-\zeta_0)\big)}}~, & \tilde{\Lambda}<0~,\\
\frac{1}{2\tilde\lambda}\sqrt{-c_1\pm4\tilde\lambda^2e^{\pm 2 \sqrt {2}\tilde\lambda(\zeta-\zeta_0)}}~,  &  \tilde{\Lambda}=0~, \\
\frac{1}{2\tilde\lambda}\sqrt{-c_1+\sqrt{\tilde{\Lambda}}\cosh{\big(2 \sqrt{2}\tilde\lambda(\zeta-\zeta_0)\big)}}\,~,  &  \tilde{\Lambda}>0~,\\
\end{array} \label{16}\right. \\
v_0(\zeta)&=\sqrt{c_1(\zeta -\zeta_0)^2-\frac{2c}{c_1}}~, \label{17}\\
v^+_{\Lambda}(\zeta)&= \left\{ \begin{array}{ll}
\frac{1}{2\lambda}\sqrt{c_1+i\sqrt{-\Lambda}\sin{\big(2 \sqrt{2}\lambda(\zeta-\zeta_0)\big)}}~, & \Lambda<0~,\\
\frac{1}{2\lambda}\sqrt{c_1\pm4\lambda^2e^{\pm 2 i\sqrt {2}\lambda(\zeta-\zeta_0)}}~,  &  \Lambda=0~, \\
\frac{1}{2\lambda}\sqrt{c_1+\sqrt{\Lambda}\sin{\big(2 \sqrt{2}\lambda(\zeta-\zeta_0)\big)}}\,~,  &  \Lambda>0~.\\
\end{array} \right. \label{18}
\end{align}
for $\lambda^2=-\tilde \lambda^2<0$, $\lambda^2=0$, and $\lambda^2>0$, corresponding to (\ref{16}), (\ref{17}), and (\ref{18}), respectively. $\tilde{\Lambda}=c_1^2-16c\tilde\lambda^2$ and $\Lambda=c_1^2+16c\lambda^2$ are discriminant quantities of the integrand in \eqref{a21}, while $c_1$ and $\zeta_0$ are arbitrary constants.

\medskip
%


It is also possible to construct a different form of the general solution for a CD-SEP equation in terms of particular solutions. 
For this, we consider the CD-SEP equation whose general solution we seek as the first equation in the following Chiellini-dissipative Ermakov-Lewis system
\begin{align}
& {\rm w}_{\zeta \zeta}+g_1({\rm w}){\rm w}_\zeta+\lambda^2{\rm w} +b {\rm w}^{-3}=0~,\quad {\rm with} \quad g_1({\rm w})=\frac{\lambda^2{\rm w}^2+b{\rm w}^{-2}}{\sqrt{-2\lambda^2 {\rm w}^4+I_{bc} {\rm w}^2+2b}}~,\label{gen-Erm}\\
& v_{\zeta\zeta}+g_2(v)v_{\zeta}+\lambda^2v+cv^{-3}=0~, \quad {\rm with} \quad g_2(v)=\frac{\lambda^2 v^2+cv^{-2}}{\sqrt{-2\lambda^2 v^4+c_1 v^2+2c}}~. \label{gen-Erm1}
 \end{align}
In the first equation, $I_{bc}$ is the Ermakov-Lewis invariant 
\begin{equation}\label{a25}
I_{bc}=-2b \Big (\frac z {\rm z} \Big)^2-2c \Big (\frac {\rm z} z \Big)^2+({\rm z}_{\zeta} z- {\rm z} z_{\zeta})^2,
\end{equation}
built from particular solutions of the non-dissipative Ermakov system of double nonlinear couplings
\begin{align}\label{ndissErm1}
 &{\rm z}_{\zeta \zeta}+\lambda^2{\rm z}+2b {\rm z}^{-3}=0~,\\
 & z_{\zeta\zeta}+\lambda^2z+2cz^{-3}=0~.
 \end{align}
If we choose particular solutions of equation \eqref{gen-Erm1} by fixing $c_1$ and $\zeta_0$ in the $v$ solutions $v$ obtained through the Abel route, then we can get general solutions of \eqref{gen-Erm} according to the following theorem.

\medskip

{\bf Theorem}.
{\em For the CD-SEP equation (\ref{gen-Erm}), with $v(\zeta)$ the particular solutions of (\ref{gen-Erm1}) and corresponding $g$'s,
the general solutions ${\rm w}(\zeta)$ are given by
\begin{eqnarray}\label{a24bis}
 \begin{array}{ll}
{\rm w}_{-}(\zeta)= q^{-}_{\Delta}(\varphi_- -\hat \varphi)
\,v_{-}(\zeta)~, & c<0~,\\
\\
{\rm w}_{0}(\zeta)=q_0(\varphi_0-\hat \varphi)
\,v_{0}(\zeta)~,  &  c=0~,\\
\\
{\rm w}_{+}(\zeta)=q^{+}_{\Delta}(\varphi_+-\hat\varphi)\,v_{+}(\zeta)~,  &  c>0~,
\end{array}
\end{eqnarray}
where the phases $\varphi(\zeta)$ are of the Milne type \cite{Milne}, i.e.,
\begin{equation}\label{ph}
\varphi_+(\zeta)=\int \frac 1{v_+^{2}}d \zeta~, \quad \varphi_0(\zeta)=\int \frac 1{v_0^{2}}d \zeta~,\quad
\varphi_-(\zeta)=\int \frac 1{v_-^{2}}d \zeta~,
\end{equation}
and $\hat \varphi$ is an initial phase. These phases have been introduced long ago by Milne in his EP approach for the Schr\"odinger equation.
The functions $q_\Delta$ are given by the following expressions:
\begin{align}\label{qtheta}
q^-_{\Delta}(\zeta)&= \left\{ \begin{array}{ll}
\frac{1}{2\sqrt{c}}
\sqrt{-I_{bc}+\sqrt{-\Delta}\sinh{\big(2 \sqrt{2c}(\varphi_{-}-\hat \varphi)\big)}}~, & \Delta<0~,\\
\frac{1}{2\sqrt{c}}\sqrt{-I_{bc}\pm4ce^{\mp 2 \sqrt {2c}(\varphi_{-}-\hat\varphi)}}~,  &  \Delta=0~,\\
\frac{1}{2\sqrt{c}}\sqrt{-I_{bc}-i\sqrt{\Delta}\sinh{\big(2 \sqrt{2c}(\varphi_{-}-\hat \varphi)\big)}}\,~,  &  \Delta>0~,\\
\end{array} \right.\\
q_0(\zeta)&=\sqrt{I_{b0}(\varphi_{0} -\hat \varphi)^2-\frac{2b}{I_{b0}}}~,\\
q^+_{\Delta}(\zeta)&= \left\{ \begin{array}{ll}
\frac{1}{2\sqrt{c}}\sqrt{-I_{bc}+\sqrt{-\Delta}\sinh{\big(2 \sqrt{2c}(\varphi_{+}-\hat\varphi)\big)}}~, & \Delta<0~,\\
\frac{1}{2\sqrt{c}}\sqrt{-I_{bc}\pm4ce^{\pm 2 \sqrt {2c}(\varphi_{+}-\hat\varphi)}}~,  &  \Delta=0~,\\
\frac{1}{2\sqrt{c}}\sqrt{-I_{bc}+\sqrt{\Delta}\cosh{\big(2 \sqrt{2c}(\varphi_{+}-\hat \varphi)\big)}}\,~,  &  \Delta>0~,\\
\end{array} \right.
\end{align}
where $\Delta=I_{bc}^2-16bc$.}



\medskip

{\bf Proof.} We start by showing that $I_{bc}$ is constant as a simple application of the connection with the Abel solution. For two equations of the CD-SEP type  with $h_b=\lambda^2{\rm z}+b{\rm z}^{-3}$ and $h_c=\lambda^2z+cz^{-3}$, for arbitrary real constants $b$ and $c$, we use the fact that ${\rm z}_{\zeta}=h_b/g_b$ and $z_{\zeta}=h_c/g_c$ to turn them into the dissipation-free form
\begin{align}\label{diss-free}
&{\rm z}_{\zeta\zeta}+2h_b({\rm z})=0~,\nonumber\\
&z_{\zeta\zeta}+2h_c(z)=0~.
\end{align}
But for this pair of EP equations it is well known that one has the Ermakov invariant $I_{bc}$ given in (\ref{a25}), see, e.g., \cite{c2}.


Then, using $\varphi=\int z^{-2}d \zeta$, and $q ={\rm z}/z$ in (\ref{a25}) we obtain the following separable equation
\begin{equation}\label{a26}
\frac{q d q}{\sqrt{2b+ I_{bc} q^2+2c q ^4}}=d \varphi~,
\end{equation}
which depending on the sign of the discriminant  $\Delta=I_{bc}^2-16bc$ of the quadratic form in the denominator is solved by cases.


Finally, since the CD-SEP format is independent of the symbols used for the unknown functions, from $q={\rm w}/v$ one gets ${\rm w}=q v$, which leads to (\ref{a24bis}).
Results similar to (\ref{a24bis}) but for the non-dissipative case can be found in a paper by Qin and Davidson \cite{qin}.

In section \ref{S6}, we will obtain explicit solutions of the type (\ref{a24bis}) for the more general case of Reid's $(2m-1)$th-order nonlinearities and write (\ref{a24bis}) as the particular case $m=2$.

\medskip
\subsection{Application using $I_{01}$ (the Courant-Snyder Invariant)}
The invariant $I_{01}$ is also known as the Courant-Snyder invariant since $I_{01}$ has been derived through the Hill equation route in their seminal work on the motion of a charged particle in alternating-gradient field configurations in accelerator physics \cite{cs58}.
The following example involving $I_{01}$ is worthwhile to show how one can compute a solution of \eqref{gen-Erm}  using the particular solution of \eqref{gen-Erm1}. Let us consider a special case 
for which $b=0$ and $c=1$, and let us choose $\lambda=1/2$. 
The dissipation-free system is the following one:
\begin{align}\label{abis}
&{\rm z}_{\zeta\zeta}+\frac{1}{2}{\rm z}=0~, \nonumber\\
&z_{\zeta\zeta}+\frac{1}{2}z+2z^{-3}=0~.
\end{align}
The particular solution to the second equation is obtained via the case $\Lambda>0$ with $c_1=1$ in \eqref{16} to give
\begin{equation}\label{v}
z(\zeta)=\sqrt{1+\sqrt5 \sin{\sqrt 2 \zeta}}~,
\end{equation}
while the first equation is a simple harmonic oscillator with particular solution
\begin{equation}\label{w}
{\rm z}(\zeta)=\sin {\frac{\zeta}{\sqrt 2}}~.
\end{equation}
Calculation of the Courant-Snyder invariant using these particular solutions gives:
\begin{equation}\label{I01}
I_{01}=-2 \Big(\frac{{\rm z}}{z}\Big)^2+({\rm z}_{\zeta} z- {\rm z} z_{\zeta})^2 = \frac 1 2~.
\end{equation}
Using the Chiellini dissipation functions, we get the following Chiellini dissipative Ermakov-Lewis system
\begin{align}\label{wv}
&{\rm w}_{\zeta\zeta}+\frac{\sqrt 2|{\rm w}|}{4\sqrt{1-{\rm w}^2}}{\rm w}_{\zeta}+\frac{1}{4}{\rm w}=0~,\\
&v_{\zeta\zeta}+\frac{\frac{v^4}{4}+v^{-2}}{\sqrt{-\frac{v^4}{2}+v^2+2}}v_{\zeta}+\frac{1}{4}v+v^{-3}=0~.
\end{align}

\begin{figure}[x!] 
\begin{center}
\resizebox*{0.35\textheight}{!}{
{\includegraphics{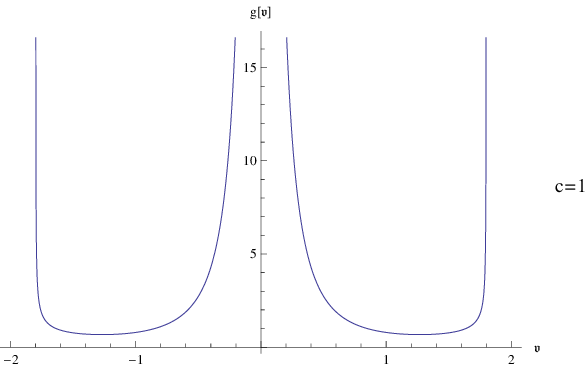}}}
\resizebox*{0.35\textheight}{!}{
{\includegraphics{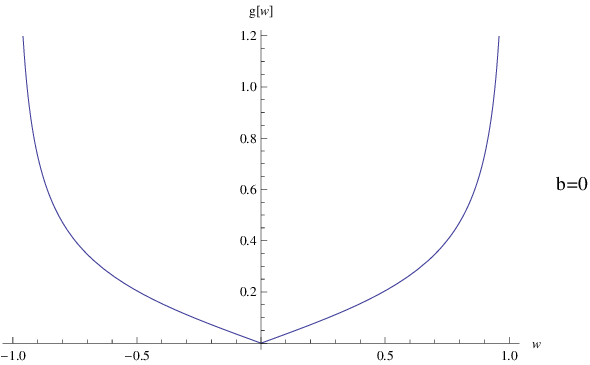}}}
\resizebox*{0.59\textheight}{!}{
{\includegraphics{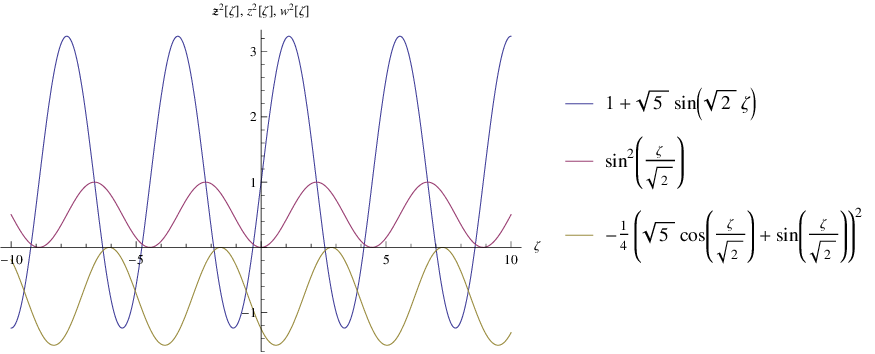}}}
\caption{\textsl{(Color online) The Courant-Snyder case $c=1,b=0, I_{01}=\frac 12$ for $\lambda=\frac 1 2$. Top: Chiellini dissipations $g(v),g({\rm w})$. Bottom: Squared solutions $z$, ${\rm z}$, and ${\rm w}$.}}
\label{Set1}
\end{center}
\end{figure}

Substituting the value of the invariant $I_{01}$ in \eqref{a26}, it yields
\begin{equation}\label{th2}
{\rm arcsinh}\, 2 q =\pm \sqrt 2(\varphi-\hat \varphi).
\end{equation}
The Milne phase is obtained using \eqref{v}, to get
\begin{equation}
\varphi=\int \frac{d \zeta}{1+ \sqrt 5 \sin \sqrt 2 \zeta}=-\frac {1}{\sqrt 2}{\rm arctanh \Big(\frac{\sqrt 5+\tan \frac{\zeta}{\sqrt 2}}{2}\Big)}~.
\end{equation}\label{eq40}
By substituting the Milne phase  in \eqref{th2}, and choosing zero initial phase, we get
\begin{equation}\label{eq41}
q= \frac{\sqrt 5 +\tan \frac{\zeta}{\sqrt 2}}{4\sqrt{1-\frac 1 4\Big(\sqrt 5 +\tan \frac{\zeta}{\sqrt 2}\Big)^2}}~.
\end{equation}
Now, the ${\rm w}$ solution is obtained from \eqref{eq41} and \eqref{v}
\begin{equation}\label{gen}
{\rm w}(\zeta)=qv= \frac{\Big(\sqrt 5 +\tan \frac{\zeta}{\sqrt 2}\Big
)\sqrt{1+ \sqrt 5 \sin \sqrt 2 \zeta}}{4\sqrt{1-\frac 1 4\Big(\sqrt 5 +\tan \frac{\zeta}{\sqrt 2}\Big)^2}}=\frac{\sqrt 5 \cos\frac{\zeta}{\sqrt 2}+\sin \frac{\zeta}{\sqrt 2}}{2 i}~.
\end{equation}
The squares of the particular solutions ${\rm z}(\zeta), z(\zeta)$ of the non-dissipative system (\ref{abis}), and the solution ${\rm w}$ are shown in Fig.~\ref{Set1}.


\section{CD-SEP equations with Reid's higher-order nonlinearities}\label{S6}

In this section, we show that the method of obtaining the general solution just described can be also applied to equations with high-order Ermakov nonlinearities and associated Chiellini dissipation.

Reid has shown in \cite{reid} that a particular solution to
\begin{equation}\label{r5}
v_{\zeta \zeta}+h(\zeta)v+\tilde{q}_m(\zeta)v^{-(2m-1)}=0
 \end{equation}
is given by
\begin{equation}\label{r4}
v(\zeta) = \left(u_1^m-\frac{c}{(m-1)W^2}u_2^m\right)^{\frac{1}{m}}
 \end{equation}
provided that $u_1$ and $u_2$ are two independent solutions of (\ref{a9}), and
\begin{equation}\label{r6}
\tilde{q}_m(\zeta) =c(u_1u_2)^{m-2}~.
 \end{equation}
Notice that (\ref{r4}) is a direct generalization of the Pinney formula (\ref{a11}).
For $h(\zeta)$ belonging to the constant triplet $(-\lambda^2,0,\lambda^2)$, the solutions of (\ref{r5}) are
\begin{eqnarray}\label{r10}
 \begin{array}{ll}
 v_{-}(\zeta)=\big(a^me^{m \lambda \zeta}-c_mb^me^{-m\lambda \zeta}\big)^{\frac{1}{m}}~, &  h(\zeta)=-\lambda^2~,\\
 \\
 v_0(\zeta)=(1-\frac{c}{m-1}\zeta^m)^{\frac{1}{m}}~, & h(\zeta)=0~,\\
 \\
v_{+}(\zeta)=\big(a^m\cos {m \lambda \zeta}-c_mb^m \sin{m\lambda \zeta}\big)^{\frac{1}{m}}~, & h(\zeta)=\lambda^2,
\end{array}
\end{eqnarray}
where
\begin{equation}\label{r11}
 c_m=\frac{c}{4\lambda^2(ab)^m(m-1)}
 \end{equation}
and $a$ and $b$ are constants determined by the initial conditions.

We are interested in a general solution to the Chiellini-dissipative equation with constant $h(\zeta)=+\lambda^2, 0, -\lambda^2$
\begin{equation}\label{r7}
{\rm w}_{\zeta \zeta}+g({\rm w}) {\rm w}_{\zeta}+h(\zeta){\rm w}+\tilde{q}_m(\zeta){\rm w}^{1-2m}=0,
 \end{equation}
via the machinery of invariants as in the previous theorem.
 For simplification, let us denote $A=a^m$, $B=-c/[4 \lambda^2 a^m (m-1)]$, and $B_0=-c/(m-1)$.
Then, by integrating $\varphi_{\mp}=\int v_{\mp}^{-2}d\zeta$ and $\varphi_{0}=\int v_{0}^{-2}d\zeta$ leads to
\begin{eqnarray}\label{r13}
 \left\{ \begin{array}{ll}
\varphi_{-}(\zeta)=\frac{Ae^{2m \lambda \zeta}+B}{2 \lambda B \big(Ae^{m \lambda \zeta}+Be^{-m\lambda \zeta}\big)^{\frac{2}{m}}}{}_2F_1\Big(1,\frac{m-1}{m};\frac{m+1}{m};-\frac{A}{B}e^{2 m \lambda \zeta}\Big)~,\\
\varphi_0(\zeta)=\zeta\, {}_2F_1\Big(\frac{1}{m},\frac{2}{m};\frac{m+1}{m};-B_0\zeta^m\Big)~, \\
\varphi_{+}(\zeta)=-\frac{\sin^{\frac{2}{m}}\big(m \lambda \zeta +\arctan\frac{A}{B}\big)\cos \big(m \lambda \zeta +\arctan\frac{A}{B}\big)}{m \lambda \big(A\cos {m \lambda \zeta}+B \sin{m\lambda \zeta}\big)^{\frac{2}{m}}}{}_2F_1\Big(\frac1 2, \frac 1 2+\frac 1 m ; \frac 3 2;\cos^2\big(m \lambda \zeta +\arctan\frac{A}{B}\big)\Big)~.
\end{array} \right.
\end{eqnarray}

Finally, the solution to (\ref{r7}) is ${\rm w}=qv$, with $q$ obtained by integrating (\ref{a26}).

\medskip

We now fix $h(\zeta)=+\frac 1 4, 0, -\frac 1 4$, together with $a=b=-c=c_1=1$, in order to obtain the following more conventional solutions that can be plotted:
\begin{eqnarray}\label{r14}
 \begin{array}{ll}
{\rm w}_{-}(\zeta,m)=\big(e^{\frac {m \zeta}{2}}+\frac{1}{m-1}e^{-\frac {m  \zeta}{2}}\big)^{\frac{1}{m}}\big(-1\mp \sqrt{3}\sinh(\sqrt{2}\varphi_{-})\big)^{\frac{1}{2}}~, & h(\zeta)=-\frac 1 4~,\\
\\
{\rm w}_{0}(\zeta,m)=\big(1+\frac{1}{m-1}\zeta^m\big)^{\frac{1}{m}}\big(\varphi_0^2-1\big)^{\frac{1}{2}}~, & h(\zeta)=0\\
\\
{\rm w}_{+}(\zeta,m)=\big(\cos{ \frac {m \zeta} 2}+\frac{1}{m-1}\sin {\frac {m \zeta}{ 2}}\big)^{\frac{1}{m}}\big(1\pm \sqrt{5}\sin(\sqrt{2}\varphi_{+})\big)^{\frac{1}{2}}~, & h(\zeta)=\frac 1 4~,
\end{array}
\end{eqnarray}

where
\begin{eqnarray}\label{r15}
 \left\{ \begin{array}{ll}
\varphi_{-}(\zeta)=\frac{(m-1)e^{m \zeta}+1}{\Big(e^{\frac {m\zeta}{2}}+\frac{e^{-\frac {m \zeta}{2}}}{m-1}\Big)^{\frac{2}{m}}}\, {}_2F_1\Big(1,\frac{m-1}{m};\frac{m+1}{m};-\left(\sqrt{m-1}e^{\frac{m \zeta}{2}}\right)^2\Big)~, \\
\varphi_0(\zeta)=\zeta\, {}_2F_1\Big(\frac{1}{m},\frac{2}{m};\frac{m+1}{m};-\frac{1}{m-1}\zeta^m\Big)~, \\
\varphi_{+}(\zeta)=\frac{2 \sin^{\frac{2}{m}}\big(\frac {m  \zeta}{2} +\arctan{(m-1)}\big)\cos \big(\frac{m\zeta}{2} +\arctan{(m-1)}\big) }{ m \big(\cos \frac {m \zeta} 2 +\frac{1}{m-1} \sin \frac {m \zeta}2 \big)^{\frac{2}{m}}} {}_2F_1\Big(\frac1 2, \frac 1 2+\frac 1 m ; \frac 3 2;\cos^2\big(\frac{m\zeta}{2} +\arctan(m-1)\big)\Big)~.
\end{array} \right.
\end{eqnarray}

\begin{figure}  [!ht]
   \centering
    \includegraphics[width= 10.15 cm, height=11 cm]{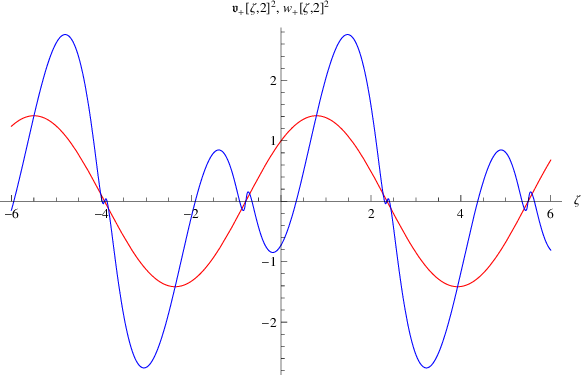}
  \caption{(Color online) The squares of the dissipative ${\rm w}_+(\zeta)$ solution (blue) and the non-dissipative $v_+(\zeta)$ solution (red) for the case $m=2$ and $a=b=-c=c_1=1,\lambda=1/2$.} 
  \label{fig-e2}
 \end{figure}

\begin{figure}  [!ht]
   \centering
    \includegraphics[width= 10.15 cm, height=11 cm]{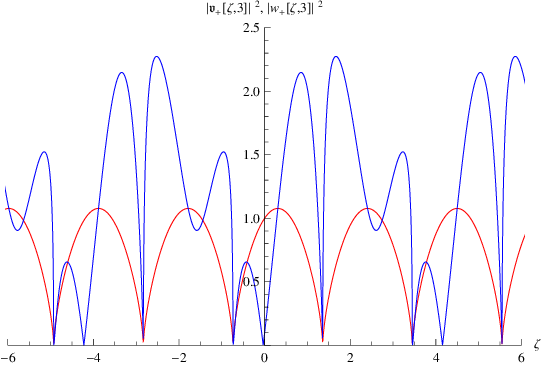}
  \caption{(Color online) The squared moduli of ${\rm w}_+(\zeta)$ and $v_+(\zeta)$ solutions for the case $m=3$ and the same color code and values of the parameters.} 
  \label{fig-e3}
 \end{figure}

\begin{figure}  [!ht]
   \centering
    \includegraphics[width= 10.15 cm, height=11 cm]{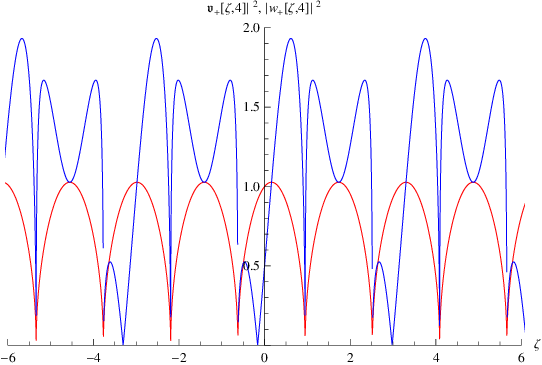}
  \caption{(Color online). The squared moduli of the dissipative solution ${\rm w}_{+}(\zeta)$ (blue) and the non-dissipative solution $v_{+}(\zeta)$ (red) for the case $m=4$ and the same values of the parameters.}   
  \label{fig-e4}
 \end{figure}

\begin{figure}[x!] 
\begin{center}
\resizebox*{0.37\textheight}{!}{
{\includegraphics{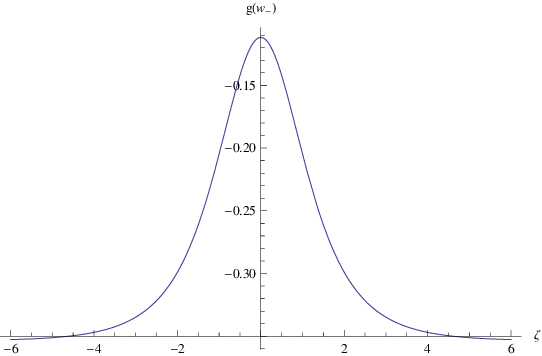}}}
\resizebox*{0.37\textheight}{!}{
{\includegraphics{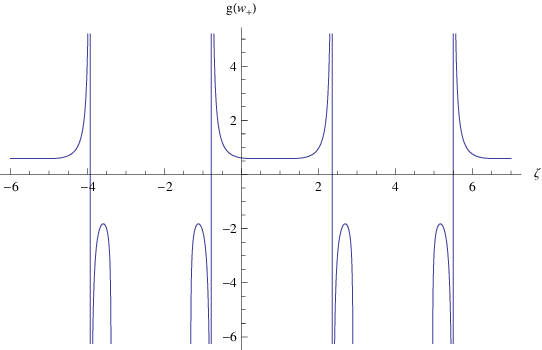}}}
\resizebox*{0.37\textheight}{!}{
{\includegraphics{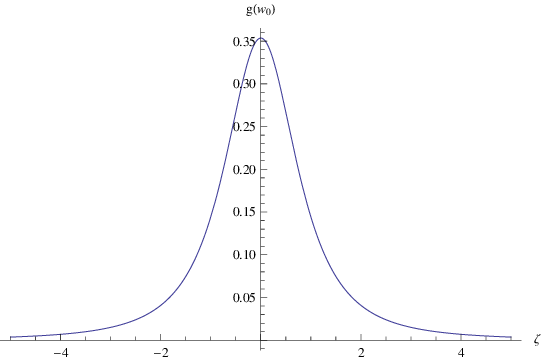}}}
\caption{\textsl{(Color online) Chiellini dissipation functions $g({\rm w}(\zeta,2))$ for: $\lambda^2=-\frac{1}{4}$, $\frac{1}{4}$, and $0$, respectively, and the other parameters as above.}}
\label{Set10}
\end{center}
\end{figure}

Note that for the case $m=2$, which is the standard dissipative SEP case, the solutions (\ref{r14}) simplify to

\begin{eqnarray}\label{r16}
 \begin{array}{ll}
{\rm w}_{-}(\zeta,2)=
\sqrt{\left(e^\zeta+e^{-\zeta}\right)\Big(-1\mp \sqrt 3 \sinh \left(\sqrt 2 \arctan e^{\zeta}\right)\Big)}~,  & h(\zeta)=-\frac 1 4~, \\
\\
{\rm w}_0(\zeta,2)=\sqrt{(\zeta^2+1)(\arctan^2 \zeta -1)}~,  & h(\zeta)=0~,\\
\\
{\rm w}_{+}(\zeta,2)=
\sqrt{(\cos\zeta+\sin \zeta ) \Big(1\pm \sqrt 5 \sin \big( \rm arctanh \frac{\cos\zeta-\sin \zeta}{\sqrt 2}\big)\Big)}~,  & h(\zeta)=\frac 1 4~.
\end{array}
\end{eqnarray}
We plot in Figs.~\ref{fig-e2}-\ref{fig-e4} the oscillating solutions ${\rm w}_{+}$ and $v_+$ for $m=2,3$, and $4$ as given in (\ref{r16}), (\ref{r14}), and (\ref{r10}), respectively.  
The interesting feature is that they are periodically pure real and pure imaginary. 
We notice also a diminishing of the amplitudes with increased order of the negative-power nonlinearity. But the most interesting feature of the dissipative solutions is that they may have larger amplitudes than the non-dissipative ones on some time intervals. This reveals the presence of gain effects. Indeed, from the plots of the functions $g({\rm w})$ in Fig.~\ref{Set10}, one can infer that this function is not always dissipative, but it also has gain intervals.

\medskip

\section{A non-linear frequency case: ${\bf h_0(v)=\Omega_0^2(v-v^2)}$}

Let us take now the case of the equation $u_{\zeta\zeta}+\omega^2(u)u=0$ with $\omega^2(u)=\pm \Omega_0^2(1-u)$, and $\Omega_0$ a real nonzero constant to obtain
\begin{equation}\label{a12p}
u_{\zeta\zeta}\pm \Omega_0^2(u-u^2) =0~.
\end{equation}
We will derive a Weierstrass $\wp$ solution of this equation below and other more particular solutions are also known.
However, since this equation is nonlinear we cannot apply the EP superposition principle to get the solution of the Ermakov-extended equation
\begin{equation}\label{a13p}
v_{\zeta\zeta}\pm \Omega_0^2(v-v^2)+cv^{-3}=0~.
\end{equation}

\medskip

On the other hand, we are able to get solutions of the corresponding CD-SEP equation:
 \begin{equation}\label{v-heqp}
 v_{\zeta \zeta}+g(v)v_{\zeta}+h(v)=0~, \qquad h(v)= \pm \Omega_0^2(v-v^2)+cv^{-3}~.
 \end{equation}
The dissipation term $g(v)$ is again obtained from $h(v)$ using Chiellini's integrability condition
for Abel's equation of the first kind \eqref{Abel2} corresponding to (\ref{v-heqp}).
\begin{equation}\label{a19p}
g(v)=\frac{\pm \Omega_0^2(v^2-v^3)+cv^{-2}}{\sqrt{k\Big(\pm\Omega_0^2(v^4-\frac{2v^5}{3}) \Big)+c_2v^2}}~.
\end{equation}

The Abel solution 
is in this case obtained from
\begin{equation}\label{a20p}
y(v)=\frac{v}{\sqrt{a_5 v^5+a_4v^4+c_2 v^2+2c}}\equiv \frac{v}{\sqrt{P_5}}~,
\end{equation}
where $a_5=\pm 4 \Omega_0^2/3$, $a_4=\mp 2 \Omega_0^2$, $c_2$ is a free integration constant, and $c$ is the Ermakov constant.

By integration, the solution to \eqref{v-heqp} is found implicitly in terms of a hyperelliptic integral of genus $(n-1)/2$, where $n=5$ is the degree of the square-rooted polynomial in (\ref{a20p})
\begin{equation}\label{a21p}
\zeta-\zeta_0=\int \frac{vdv}{\sqrt{P_5}}~.
\end{equation}
This Jacobi inversion problem can be solved in terms of hyperelliptic functions of two variables with periods which are two-by-two matrices \cite{Enolski11}.

When the Ermakov constant is zero \eqref{a13p} becomes \eqref{a12p}, while \eqref{a21p} becomes the elliptic equation
\begin{equation}\label{a22p}
\zeta-\zeta_0=\int \frac{du}{\sqrt{{a_5 u^3+a_4u^2+c_2}}}~,
\end{equation}
with solutions in terms of Weierstrass $\wp$ functions.
Because \eqref{a22p} is not in standard form, we use a linear transformation $u=a t +b$, where $a=\sqrt[3]{4/a_5}$ and $b=-1/2\pm\sqrt{1/4-a_4\sqrt[3]{4 a_5^2}/3}$, to turn it into
\begin{equation}\label{a23p}
\zeta-\zeta_0=a\int \frac{dt}{\sqrt{{4t^3-g_2t-g_3}}}~,
\end{equation}
with solution
\begin{equation}\label{eq60}
u(\zeta)=a\wp^{-1}\Big(\frac{\zeta-\zeta_0}{a},g_2,g_3\Big)+b,
\end{equation}
with $a=\sqrt[3]{3/\Omega_0^2}$, $b=-1/2\pm\sqrt{1/4+(8/9)\Omega_0^3\sqrt[3]{3\Omega_0}}$.

The invariants of the $\wp$ function are given by
\begin{eqnarray}\label{a14ppp}
 \begin{array}{ll}
g_2&=-3 a b^2 a_5-2 a b a_4~,\\
\\
g_3&=b^3 a_5+b^2 a _4+c_2~,
\end{array}
\end{eqnarray}
and they only depend on the frequency parameter $\Omega_0$, and the integration constant $c_2$.
Depending upon the signs of $g_2$ and $g_3$, together with the discriminant $\Delta_\wp=g_2^3-27g_3^2$, \eqref{eq60} can be reduced to elementary periodic or hyperbolic functions, see \cite{ww}.

\section{Conclusion}\label{S-Concl}

A class of dissipative Ermakov-Pinney equations, either with standard $m=2$ or higher-order ($m>2$) Reid nonlinearities, and with nonlinear dissipation of the Chiellini type has been introduced. The general solutions are obtained directly through the Abel equation route, and also using the dynamic invariant of Ermakov systems of this type of equations and the particular solution of one member of the pair, both in the standard Ermakov case and in the case of any Reid nonlinearity. The technique based on Abel's equation we used here can be applied only to the constant frequency systems and cannot be directly generalized to the case of time-dependent oscillators. This is due to the fact that the Chiellini integrability condition which plays a central role in obtaining the results does not apply to the cases with an explicit dependence on the independent variable.
On the other hand, if Reid higher-order inverse power nonlinearities or other type of additional nonlinearities are introduced, one can still obtain integrable Chiellini-dissipative equations. We have discussed Reid examples of any order $m$ and also an example with an additional quadratic nonlinearity which led to a hyperelliptic case. Another remarkable aspect is that the Chiellini nonlinear dissipative function is in many cases a dissipation-gain function. If it can be engineered, it might have interesting applications in the propagation of laser beams and pulses in nonlinear optics and accelerators.

\bigskip
\bigskip


{\bf Appendix: The reduced CD-SEP equation}\\

A very interesting case occurs if we take the nonlinear coupling $c=0$ in the CD-SEP equation \eqref{v-heq}. Then, we obtain an equation that we call the reduced CD-SEP equation, which is nonlinear only because of the damping of Chiellini type:
 \begin{equation}\label{v-hred}
 u_{\zeta \zeta}+g_r(u)u_{\zeta}+\lambda^2u=0~, \qquad g_r(u)=\frac{\lambda^2 u}{\sqrt{c_1-2\lambda^2u^2}}~.
 \end{equation}
Despite being nonlinear, this equation has for $\lambda^2>0$ the linear harmonic solutions $u_{1r}=(\sqrt{c_1}/\sqrt{2}\lambda)\sin \sqrt{2}\lambda (\zeta-\zeta_0)$ and $u_{2r}=(\sqrt{c_1}/\sqrt{2}\lambda)\cos \sqrt{2}\lambda (\zeta-\zeta_0)$ as if the nonlinear dissipation does not act at all and if judged according to its solutions the equation \eqref{v-hred} is linear. This can be checked by direct substitution. 
 The only feature introduced by the reduced nonlinear Chiellini dissipation is that the amplitudes of the harmonic modes are inverse proportional to the frequency, which in fact is an EP fingerprint. Thus, one can also obtain solutions of the reduced equation \eqref{v-hred} from the solutions (16)-(18)
 by taking $c=0$, which also implies $\Lambda=\tilde{\Lambda}=c_1^2>0$: 
  \begin{eqnarray}\label{rs10b}
 \begin{array}{ll}
u_{-}(\zeta)=\frac{\sqrt c_1}{\sqrt {2} \tilde{\lambda}}\sqrt{-1+\cosh\big(2\sqrt 2 \tilde \lambda(\zeta-\zeta_0)\big)}
\equiv \frac{\sqrt c_1}{\tilde{\lambda}}|\sinh \sqrt {2}\tilde \lambda(\zeta-\zeta_0)|
~, & \, \lambda^2=-\tilde{\lambda}^2<0~,\\
\\
u_0(\zeta)=\sqrt{c_1}(\zeta-\zeta_0)~, & \,\lambda^2=0~,\\
\\
u_{+}(\zeta)=~\frac{\sqrt c_1}{\sqrt {2} \lambda}\sqrt{1+\sin \big(2\sqrt{2} \lambda(\zeta-\zeta_0)\big)}
\equiv \frac{\sqrt c_1}{\sqrt {2} \lambda}|\sin \sqrt {2}\lambda(\zeta-\zeta_0)+ \cos \sqrt {2}\lambda(\zeta-\zeta_0)| , & \, \lambda^2>0.
\end{array}
\end{eqnarray}
Notice also that the integration constant $c_1$ should not be zero because it occurs in the amplitude of the reduced harmonic modes.

\medskip

Moreover, solutions of the type (\ref{a24bis}) can be also written for the reduced Chiellini-dissipative equations. If one considers two such equations with different $c_1$'s, say $c_1$ and $c'_1$, but the same $\lambda^2$, then their Ermakov-Lewis invariant is
$I_{00}=({\rm w}_{\zeta} v- {\rm w} v_{\zeta})^2$ and since the two constants $c_1$ figure only in the amplitudes, we have $I_{00}=c_1c'_1$.

\renewcommand{\baselinestretch}{1.0}

\renewcommand{\baselinestretch}{1.0}

\end{document}